\documentclass[submission]{eptcs}
\providecommand{\event}{SecCo 2010}

\usepackage[utf8]{inputenc}
\usepackage{times}
\usepackage{graphicx}
\usepackage{amssymb, amsbsy, amsmath, txfonts, calc, mdwlist, stmaryrd,verbatim, esint}
\usepackage{graphics,bussproofs,program, proof}

\newcommand{\secret}{\mathsf{S}}
\newcommand{\knows}{\mathbb{K}}

\newtheorem{theorem}{Theorem}[section]

\newtheorem{proposition}[theorem]{Proposition}

\newtheorem{lemma}[theorem]{Lemma}
\newtheorem{definition}[theorem]{Definition}

\def\titlerunning{A Spatial-Epistemic Logic for Reasoning about Security Protocols}
\def\authorrunning{B. Toninho, L. Caires}

\title{A Spatial-Epistemic Logic for \\ Reasoning about Security Protocols}
\author{Bernardo Toninho 
\institute{CITI and Faculdade de Ciências e Tecnologia, Universidade Nova de
  Lisboa}
\institute{Carnegie Mellon University, Pittsburgh PA, USA}
\email{Btoninho@cs.cmu.edu}
\and 
Luís Caires
\institute{CITI and Faculdade de Ciências e Tecnologia, Universidade
  Nova de Lisboa}
\email{Luis.Caires@fct.unl.pt}}

\begin{document}

\maketitle
\begin{abstract}
  Reasoning about security properties involves reasoning about
  \emph{where} the information of a system is
  located, and how it evolves over time.
While most security analysis techniques need to cope with some notions
of information locality and knowledge propagation, usually they
do not provide a general language for expressing 
arbitrary properties involving local knowledge and knowledge transfer.
 Building on this observation,
  we introduce a framework for security protocol analysis based on
  dynamic spatial logic specifications. Our computational
  model is a variant of existing $\pi$-calculi, while specifications
  are expressed in a dynamic spatial logic extended with an epistemic
  operator. We present the syntax and semantics of the model
  and logic, and discuss the expressiveness of the approach, showing
  it complete for passive attackers. We also
  prove that generic Dolev-Yao attackers may be mechanically determined for any
  deterministic finite
  protocol, and discuss how this result may be used
  to reason about security properties of open systems.  We also
  present a model-checking algorithm for our logic, which has been
  implemented as an extension to the SLMC system. 
\end{abstract}

\section{Introduction}
Among the several artifacts in the field of computer security,
security protocols are indubitably a fundamental subject of study and
research \cite{891726,358740}.  
Security protocols serve a
variety of purposes, ranging from secrecy and authentication to
forward secrecy and deniable encryption. A common trait of 
these protocols is their notoriously
difficult design, which often leads to unforeseen
vulnerabilities.

Therefore, it becomes essential to develop techniques
that ensure the correctness of protocols, with respect to some
specification of the properties they aim to establish. A wide
range of language-based techniques have been proposed to analyze
protocols and their correctness, such as type systems, process
calculi or static analysis
\cite{266432,77649,AbadiFournet01:0} which in many cases
result in successful tools \cite{BlanchetCSFW01,Armando05,Lowe98casper:a,Cr2008Scyther}. 

In this paper we propose a framework for protocol analysis based
on process calculus models and logic specifications. While the usage
of process calculi and logic in this context is not new
\cite{1431531,CDK-forte09,AbadiFournet01:0}, our approach stems from
the fact that many interesting properties of such systems are often 
a function of what information the several parts of a system may or may not
obtain. While other frameworks (e.g., Avispa \cite{Armando05} and
Casper \cite{Lowe98casper:a}) allow one
to efficiently verify a wide range
of interesting security properties, these are not usually stated in
this high-level knowledge oriented approach.

Our contribution consists of a dynamic spatial epistemic logic that allows
reasoning about systems (modelled in a variant of the applied
$\pi$-calculus \cite{AbadiFournet01:0}) at three levels: the \emph{dynamics} of
systems and subsystems, the \emph{spatial} arrangement of systems and
subsystems, and the \emph{knowledge} (the obtainable information) of
systems and subsystems. The goal is to produce an expressive property
language with which we can reason about a protocol by separating
it into its different agents (malicious and otherwise), and then reason about the knowledge
they can obtain and how it can evolve over time. This enables us to
express interesting security properties in a very direct 
way (eg. agents $P$ and $Q$ can obtain value $v$, while agents $A$ and
$S$ cannot).
To clarify our approach, consider the example of Fig.~\ref{fig:mot_ex}.
\begin{figure}[tp]
\begin{center}
{\scriptsize
\begin{verbatim}

S(pkp,pkq, sks, pks) = c?(h).select { [pkp = getpk(h)].c!(enc_as(pkp,sks)).S(pkp,pkq) ; 
                                      [pkq = getpk(h)].c!(enc_as(pkq, sks)).S(pkp,pkq) };

defproc P(skp,hostQ,pks) = c!(hostQ).c?(m).let pkQ = dec_as(m,pks) in 
                            new sK in c!(enc_as(sK,pkQ)).c!(enc(v,sK)).ok!(v);
         
Q(skq) = c?(m1).let sK = dec_as(m1,skq) in c?(m2).let val = dec(m2,sK) in ok!(val);
           
defproc Sys = new skp, skq in let pkp = pk(skp) in let pkq = pk(skq)
                           in let hP = host(pkp) in let hQ = host(pkq)
                           in (S(pkp,pkq) | P(skp,hQ) | Q(skq));

World = Sys | Attacker(Sys);

prop pqK = eventually (knows v | knows v | not (knows v))
           and always (2 | not (knows v));

check World |= pqK;

* Process World satisfies the formula pqK *
\end{verbatim}
}
\caption{A Motivating Example}\label{fig:mot_ex}
\vspace{-0.2in}
\end{center}
\end{figure}

We have a system \verb=Sys= composed of three processes: 
$P$, $Q$ and a key distribution server $S$.
$P$ wishes to exchange a value $v$ with $Q$. To do so, he requests
$Q$'s public key from $S$, which $S$ emits in a signed message. 
$P$ then uses it to encrypt a generated symmetric
session key and sends the key to $Q$. Afterwards $P$ will send $v$ 
encrypted with the session key and terminate. $Q$ will receive the
message, decrypt it to obtain $v$ and terminate. We further model 
the system running with a malicious agent, defined through the
primitive \verb=Attacker(Sys)=. This agent consists of a Dolev-Yao 
attacker which we discuss in Section~\ref{sect:attackers}.

For this protocol to be correct, it must be the case that
the malicious agent interacting with the system can never know
$v$. Consider, however, a slightly stronger property: $P$ and $Q$ want 
to exchange $v$ securely (with
respect to the malicious agent) but they also do not 
completely trust the server $S$. They trust it to at least distribute the 
appropriate keys,
but want some assurances that even though $S$ operates according
to protocol, it doesn't obtain the value $v$ by observing the data 
exchanges between $P$ and $Q$. 

This property, while not impossible to state in other frameworks,
would usually require some sort of \emph{ad-hoc} modification to the
model (e.g, internalizing the server in an attacker, which seems like
an indirect strategy at best and may not necessarily yield the
correct model). 
In our framework, the property can be directly stated
by combining our epistemic and spatial operators. A formula that 
reflects such a
property is \verb=pqK=: first we state that the system can evolve
to a configuration where two of its subsystems ($P$ and $Q$) know 
$v$, but the remaining parts of the system do not. This illustrates
the expressiveness of the logic in terms of reasoning about the 
knowledge of several parts of the complex system. Secondly,
we state that throughout \emph{all} executions of the system, 
a part of it will never know $v$ (\verb=2= indicates that there
must be two agents running with the part that does not know $v$ -- a
precise definition is given in Section~\ref{sect:logic}).
By combining spatial reasoning with
epistemic reasoning, we can state rich properties of the
knowledge of agents (and groups of agents) -- both adversaries and principals -- within a
complex system, and how they can share or restrict that knowledge over time.

While our framework is aimed at reasoning about closed systems, 
meaningful analyses of security protocols must necessarily consider attackers.
Traditionally, attackers are modelled by an adversarial environment
which interacts with the protocol. In our closed system approach, 
we develop a way internalizing an
arbitrary attacker within a closed system by \emph{automatically} (or
semi-automatically) deriving a process representation of such an
attacker. This representation makes use of special primitives built
into the process calculus to greatly simplify the actual modelling of
the attacker, to the extent that even if the attacker generation is
done semi-automatically, it \emph{never} requires us to actually
encode specific attacks. In this work, we show that it is
possible to \emph{automatically} derive an attacker (behaving as a
Dolev-Yao adversarial environment) for any
finite protocol. To fully automate our technique (at
the implementation level),
further work is needed (as discussed in Section~\ref{sect:attackers});
the focus of the current paper being essentially on the expressiveness
issues.  In any case, we already provide tool support for arbitrary
passive attackers and for \emph{bounded} Dolev-Yao attackers (where
the bound concerns the size of generated messages); this technique can
already be used to automatically find attacks, eg., as illustrated in
the example of Section~\ref{sect:correspondence}.

The technical contributions of this work are as follows.  We
develop a process calculus model for security protocols (Section
\ref{sect:model}), inspired in existing $\pi$-calculi, supporting
explicit modeling of adversarial agents, at an adequate level of
abstraction. We introduce a new dynamic spatial epistemic logic
(Section \ref{sect:logic}), oriented for reasoning about spatial
distribution of information.  We develop a logic-based theory of
knowledge deduction (Section \ref{sect:proof}) for our
models, proved sound, complete and decidable. 
This presentation was
used in our model-checking algorithms. We discuss attacker
representations (Section \ref{sect:attackers}), and how it is possible
to produce a generic Dolev-Yao attacker for finite protocols. We also
show how to model and verify correspondence assertions
(Section~\ref{sect:correspondence}) in our framework. Finally, we
implemented a model-checking algorithm for the logic as an
extension to the SLMC tool, producing the first proof of concept tool aimed at security
protocol analysis using spatial logic model checking. The proofs
of our technical results are detailed in \cite{toninhocaires09:0}.

\section{Process Model}\label{sect:pmodel}

In this section we introduce our process model, starting with some
preliminary notions on terms and equational theory and then
introducing our process calculus.

\subsection{Terms and Equational Theories}\label{sect:terms}

Data exchanged by processes is modeled by terms of a term algebra.  
In order to capture cryptographic operations and data
structuring, we will consider term algebras with 
equational theories (cf.~\cite{AbadiFournet01:0}).

We assume an infinite set of variables ranged over by
$x,y,z$, an infinite set of names $\Lambda$ ranged over by $m,n$ 
and range over terms with $s,t,v$. 
Terms are defined from names and variables by applying function
symbols. We thus
consider a given term algebra to be defined from a signature $\Sigma$
and an equational theory $E$ that defines the ``semantics''
of the function symbols in $\Sigma$.
An equational theory is a congruence relation defined by a set of
equations of form $t = s$.

In certain circumstances, an equational theory may give rise to a set
of rewrite rules by orienting each equation to produce the rule $t
\rightarrow s$, in such a way that two terms are equal modulo $E$
whenever that have a common reduct under rewriting.  This is the case
of subterm convergent equational theories \cite{AbadiCortier06:0},
which are the ones that we will focus on in this work (other
equational theories, such as AC theories, can also be applied in this
fashion, however with a slightly different formal treatment as
detailed in \cite{AbadiCortier06:0}). A subterm 
convergent system is a convergent rewrite system in which in every 
rewrite rule
the right-hand side is a proper subterm of the left-hand side. In this
paper, we will assume a general rewrite theory $\mathcal{R}$ subject
to the conditions above.
Given a rewrite rule $t \rightarrow s$, we call the outermost function
symbol in $t$ a \emph{destructor}, since the application of the rule
may open the internal structure of inner terms in $t$ to produce the
term $s$.  We classify the remaining function symbols, that never
occur as a destructor, as \emph{constructors}. For example, for
signature $\Sigma \triangleq \{ \mathtt{enc}/2 \mbox{;} \mathtt{dec}/2
\}$ and equational theory $E \triangleq \{
\mathtt{dec}(\mathtt{enc}(x,y),y) = x \}$, $\mathtt{dec}$ is a
destructor and $\mathtt{enc}$ a constructor. We range over
constructors with $f$ and destructors with $\delta$.

We denote the set of names of a term $T$ by $names(T)$ and the depth
of a term as $\origbar T \origbar$ (the depth is the length of the
longest path in the tree representation of the term). We state that a term is ground
if it does not contain variables.
We denote by $=_E$ the usual congruence relation induced by the 
set of equations $E$ (which can be decided through term rewriting since
$R$ is convergent).
We write $\mathfrak{F}(\psi)$ for the DY (Dolev-Yao) equational
closure of a set of terms $\psi$, that is, the set of all values (destructor-free terms)
generated by terms of $\psi$ through function application, modulo the
equational theory.
This closure represents all possible information that may be produced
from a set of terms while following the rules of the equational
theory, which if we interpret a set of terms as a set of messages, is
the usual notion of knowledge from the Dolev-Yao model.
\begin{definition}[Equational Closure]
  Given a rewrite theory $\mathcal{R}$, the DY equational closure of a
  set of terms $\psi$, noted $\mathfrak{F}(\psi)$, is the least set of
  terms such that: {\small
\begin{enumerate}
\item $\psi \subseteq \mathfrak{F}(\psi)$
\item $\forall f \in \Sigma . \mbox{ if }  f \mbox{ a constructor and } t_{1},\dots, t_{k} \in \mathfrak{F}(\psi) \mbox{ then }  f(t_{1},\dots,t_{k}) \in \mathfrak{F}(\psi)$
\item $\forall \delta \in \Sigma . \mbox{ if } \delta \mbox{ a destructor and }
  t_{1}, \dots, t_{k} \in \mathfrak{F}(\psi)
\mbox{ and } \delta(t_{1}, \dots, t_{k}) \rightarrow t' \mbox{ then } t' \in \mathfrak{F}(\psi)$
\end{enumerate}}
\end{definition}
When interpreting the DY equational closure of a set of terms as
obtainable knowledge, we can state knowledge derivation through term
derivation.
\begin{definition}{\textbf{(Knowledge Derivation).}}
  Given sets of terms $\psi$ and $\phi$, we say that $\phi$ may be
  derived from $\psi$ (written $\psi \models \phi$) if and only if
  $\phi \subseteq \mathfrak{F}(\psi) $.
\end{definition}
The general idea is that one may can derive a piece of information if
it can be generated by combining pieces of information using the rules
of the equational theory.  Given these basic notions relative to
terms, equational theories, and knowledge derivation, we may now
present our process calculus model.

\subsection{Process Calculus}\label{sect:model}

It is known that the high level of abstraction of the $\pi$-calculus,
convenient from a foundational perspective, is not suitable for
modeling cryptographic techniques as necessary for analyzing security
protocols.

We therefore adopt an extension to the $\pi$-calculus that extends the base
values of the language with functional terms (cf. Section
\ref{sect:terms}), that can be seen as a fragment of the Applied
$\pi$-calculus \cite{AbadiFournet01:0}.
We choose this calculus over the applied $\pi$-calculus mainly for
simplicity reasons, not requiring active substitutions nor frames
given that our goal is to use our logic to observe terms.

We model cryptographic operations by defining such operations in a
term algebra.  The calculus is thus aimed at the explicit modeling of
agents involved in security protocols, both principals and
adversaries. Principals are modeled standardly, using terms to model
cryptographic terms. Adversaries are modeled as processes (cf. Section
\ref{sect:attackers}) using the attacker output prefix - a
non-deterministic output of terms that can be generated from known
values, which enables reasoning directly about attacker knowledge
using our logic.

\begin{definition}[Processes]
Given a signature $\displaystyle\Sigma$, an infinite set of names ranged over by $m,n$,
and an infinite set of variables ranged over by $x,y,z$,
the set of processes $(P,Q)$, of actions $\alpha$ and of terms $T$ are defined in Fig. \ref{fig:pi_syntax}.
\begin{figure}[tp]
\begin{center}
$$
{\small
\begin{array}{llll}
P,Q & \Coloneqq & \mathbf{0} & \mbox{(Null Process)} \\
   & \;\;\;\origbar & P\;\origbar\; Q & \mbox{(Parallel Composition)}\\
   & \;\;\;\origbar & (\boldsymbol{\nu} n)P & \mbox{(Name Restriction)}\\
   & \;\;\;\origbar & \alpha.P & \mbox{(Action Prefix)}\\
   & \;\;\;\origbar & P+Q & \mbox{(Choice)}\\
   & \;\;\;\origbar & \mbox{let } x = T \mbox{ in } P & \mbox{(Let Construct)} \\
\end{array}\quad
\begin{array}{llll}
\alpha & \Coloneqq & m(x) & \mbox{(Input)}\\
   & \;\;\;\origbar & m\langle T \rangle & \mbox{(Output)}\\
   & \;\;\;\origbar & m\langle * \rangle & \mbox{(Attacker Output)}\\
   & \;\;\;\origbar & [ T_1 = T_2 ] & \mbox{(Test)}\\
   \\
T & \Coloneqq & n & \mbox{(Name)} \\
   & \;\;\;\origbar & x & \mbox{(Variable)}\\
    & \;\;\;\origbar & f(T_{1}, \dots , T_{a}) & \mbox{(Function)} \\
    \\
\end{array}}
\vspace{-0.3in}
$$
\caption{Process Calculus Syntax}\label{fig:pi_syntax}
\vspace{-0.3in}
\end{center}
\end{figure}
\end{definition}

Before introducing the semantics of our calculus, we present some
definitions that pertain to obtaining the \emph{relevant terms} of a process 
that are necessary for our semantics.

A destructor function symbol denotes computation at the term level. If such computations
are valid (under the equational theory), then the term containing the destructor can be rewritten as one that only has constructors. On the other hand,
if such a term cannot be reduced (e.g
$\mathtt{dec}(\mathtt{enc}(m,k_1),k_2)$), it has no interesting
meaning and has no place being communicated.
To obtain the values (destructor free normal forms) of a process, we define a relation $\vdash_{k}$ that extracts the set of values $\psi$ that occur in a process ($P \vdash_k \psi$).
However, some care is needed in the definition of $\vdash_{k}$ since a term may contain bound names or variables. For instance, in the process
$a(x).a\langle \mathtt{dec}(x,k)\rangle.\mathbf{0}$, the term $\mathtt{dec}(x,k)$ is not a proper value since it contains the variable $x$. In these situations,
our extraction has to be such that it will produce a set containing $k$ but not $x$ (nor $\mathtt{dec}(x,k)$). Similarly, when we consider the terms
that are to be the object of our attacker output, while it is true that outputting a term containing a variable would be senseless, it is correct to
output a term that contains a restricted name, even though the
attacker may not be able to use the name in other messages.

To take all this into account, we define a procedure $sub$ that
extracts the \emph{relevant subterms} (not containing variables or
destructors) of a term, 
and a procedure $\uparrow$ used to eliminate terms with restricted names. 
\begin{definition}[Relevant Subterms]
Given a term $M$ we define the set of its relevant subterms, written $sub(M)$, by the rules of Fig. \ref{fig:sub}.
\begin{figure}[tp]
\begin{center}
{\em
$$
sub(\delta(t_1,\dots,t_n)) \triangleq sub(t_1) \cup \dots \cup sub(t_n)\quad
\frac{\mbox{n is a name}}
{sub(n) \triangleq n}
\quad
\frac{\mbox{x is a variable}}
{sub(x) \triangleq \emptyset}
$$
$$
\frac{\not\exists t_i\mbox{ with a variable or a destructor}}
{sub(f(t_1,\dots,t_n)) \triangleq f(t_1,\dots,t_n)}
\quad
\AxiomC{$\exists t_i$ with a variable or a destructor}
\UnaryInfC{$sub(f(t_1,\dots,t_n)) \triangleq  sub(t_1) \cup \dots \cup sub(t_n)$}\DisplayProof
$$}
\vspace{-0.2in}
\caption{Relevant Subterms}\label{fig:sub}
\end{center}
\vspace{-0.2in}
\end{figure}
\end{definition}
\begin{definition}[Name Occurrence Term Removal]
We define the removal of terms from a set $\psi$ in which the name $x$ occurs, $\psi \uparrow x$, as:
$
\quad\psi \uparrow x \triangleq \{ t \;\origbar \; t \in \psi : x \not\in names(t) \}
$.
\end{definition}
\begin{definition}[Relevant Term Extraction]\label{def:term_extraction}
Given a process $P$, the set $\psi$ of relevant terms of $P$, written $P \vdash_k \psi$, is defined by the rules of Fig. \ref{fig:term_extraction}.
\begin{figure}[tp]
\begin{center}
\vspace{-0.25in}
{\small
$$\displaystyle
\frac{P \vdash_{k} \varphi \;\;\;\; Q \vdash_{k} \psi}{\;\;P+Q \vdash_{k} (\varphi \cup \psi) \;\;}\quad
\frac{P \vdash_{k} \varphi \;\;\;\; Q \vdash_{k} \psi}{\;\; P\;\origbar\; Q \vdash_{k} (\varphi \cup \psi) \;\;}\quad
\frac{P \vdash_{k} \varphi}{\;\;n(x).P \vdash_{k} \varphi\;\;} \quad
\frac{P\{ n \leftarrow M \} \vdash_{k} \varphi}{\;\;\mbox{let } n = M \mbox{ in } P \vdash_{k} \varphi \cup sub(M) \;\;}
$$
$$\displaystyle
\frac{}{\;\;\mathbf{0} \vdash_{k} \emptyset \;\;}\quad
\frac{P \vdash_{k} \varphi}{\;\;x\langle M \rangle.P \vdash_{k} \varphi \cup sub(M)\;\;}\quad
\frac{P \vdash_{k} \varphi}{\;\; (\boldsymbol{\nu}n)P \vdash_{k}
  \varphi \uparrow n \;\;}\quad
\frac{P \vdash_{k} \varphi}{\;\;[M=N].P \vdash_{k} \varphi \cup sub(M) \cup sub(N) \;\;}
$$
}
\vspace{-0.15in}
\caption{Relevant Term Extraction}\label{fig:term_extraction}
\vspace{-0.3in}
\end{center}
\end{figure}
\end{definition}
For our attacker output we collect all ground terms that occur in the process, which we denote by $gt(P)$.

The semantics of our calculus are defined standardly, modulo $\alpha$-conversion of bound names and variables, by a structural congruence relation, labelled transition and reduction, 
as follows. We denote by $fn(P)$ and $fv(P)$ the set of free-names and
free-variables of process $P$, respectively. 
\begin{definition}[Structural Congruence]
Structural congruence $\equiv$ is the least congruence relation on processes defined by the rules of
Fig. \ref{fig:equiv}.
\begin{figure}[tp]
\begin{center}
{\small
$$
\begin{array}{ll}
n \not\in fn(P) \cup fv(P) \Rightarrow P \;\origbar\; (\boldsymbol{\nu}n)Q \equiv (\boldsymbol{\nu}n)(P \;\origbar\; Q) & \mbox{} \\
(\boldsymbol{\nu}n) \mathbf{0} \equiv \mathbf{0} & \mbox{}\\
(\boldsymbol{\nu}n)(\boldsymbol{\nu}m)P \equiv (\boldsymbol{\nu}m)(\boldsymbol{\nu}n)P & \mbox{} \\
M =_{E} M' \Rightarrow \mbox{let } x = M \mbox{ in } P \equiv \mbox{let } x = M' \mbox{ in } P &  \mbox{}\\
M =_{E} M' \Rightarrow m\langle M \rangle .P \equiv m\langle M' \rangle .P & \mbox{} \\
M_{1} =_{E} M'_{1} \Rightarrow [M_{1} = M_{2}].P \equiv [M'_{1} = M_{2}].P & \mbox{}
\end{array} \quad
\begin{array}{ll}
P \;\origbar\; \mathbf{0} \equiv P & \mbox{} \\
P \;\origbar\; Q \equiv Q \;\origbar\; P & \mbox{} \\
P \;\origbar\; (Q \;\origbar\; R) \equiv (P \;\origbar\; Q) \;\origbar\; R & \mbox{} \\
P + Q \equiv Q + P & \mbox{} \\
P + (Q + R) \equiv (P+Q) + R & \mbox{}\\
\left[ M_{1} = M_{2} \right].P \equiv \left[ M_{2} = M_{1} \right].P & \mbox{}\\
\end{array}
$$}
\vspace{-0.25in}
\caption{Structural Congruence}\label{fig:equiv}
\vspace{-0.3in}
\end{center}
\end{figure}
\end{definition}
We augment the standard structural congruence laws of the $\pi$-calculus with rules that equate processes modulo the equality $=_E$ of
the equational theory. These laws are essential in our semantics because they allow us to block processes performing actions that use terms
that are not values (i.e. terms that contain destructors). 

Our semantics, which we now present, capture these \emph{destructor freedom} conditions. If a process is attempting to use a term
that contains a destructor, we use structural congruence to rewrite the term destructor-free and reduction proceeds. If the term cannot
be rewritten destructor-free, reduction halts. These restrictions
ensure that all received terms are actual values, and not some
arbitrary erroneous term. Note the semantics
of our attacker output, expressed in the \emph{Attacker} rule, that
enable the output to emit any message that can be generated by the
process, given its ground terms and some fresh values.

\begin{definition}[Reduction Semantics]
The reduction relation $P \longrightarrow Q$ over closed processes is defined as the least relation closed under the rules of Fig. \ref{fig:reduc_sem}.
\begin{figure}[tp]
\begin{center}
{\em\small
$$\displaystyle
\frac{\mbox{\small{M is destructor-free}}}{\;\mbox{let } x = M \mbox{ in } P \longrightarrow P\{x \leftarrow M\}\;}\mbox{\small(Let)}\quad
\frac{\mbox{\small{M is destructor-free}}}{\;n\langle M \rangle.P+R\; \origbar \;n(x).Q+S \longrightarrow P\;\origbar\; Q\{x \leftarrow M\}\;}\mbox{\small(Sync)}
$$
$$\displaystyle
\frac{M_1 \mbox{\small{ and }} M_2 \mbox{\small{ are destructor-free}}\quad M_1 =_E M_2}{\; \left[ M_{1} = M_{2} \right] .P \longrightarrow P \;}\mbox{\small(Test)}\quad\quad
\frac{P \longrightarrow Q}{\;P\;\origbar\; R \longrightarrow
  Q\;\origbar\; R\;}\mbox{\small(Par)}\quad\quad
\frac{P \longrightarrow Q}{\;(\boldsymbol{\nu} n)P \longrightarrow (\boldsymbol{\nu} n)Q\;}\mbox{\small(Scope)}
$$
$$\displaystyle
\frac{\;P \equiv P' \;\;\; P' \longrightarrow Q' \;\;\; Q' \equiv
  Q\;}{\;P \longrightarrow Q\;}\mbox{\small(Cong)}
\quad\quad
\frac{M\in \mathfrak{F}(ct(Q) \cup \bar{n})\;\;\;\bar{n} \mbox{  fresh}}
{\;c(x).P + R \;\origbar\; c\langle * \rangle.Q + S \longrightarrow (\boldsymbol{\nu}\bar{n})(P\{x \leftarrow M \} \;\origbar\; Q)\;}\mbox{\small{(Attacker)}}
$$}
\vspace{-0.2in}
\caption{Reduction Semantics}\label{fig:reduc_sem}
\vspace{-0.3in}
\end{center}
\end{figure}
\end{definition}

\begin{definition}[Labelled Transition Semantics]
The labelled transition relation $P \stackrel{\alpha}{\longrightarrow} Q$ is the least relation on closed processes closed under the rules of Fig. \ref{fig:lts}.
\begin{figure}[tp]
\begin{center}
\small{
$$
\AxiomC{$P \longrightarrow Q$}
\RightLabel{\small{(Tau)}}
\UnaryInfC{$P \stackrel{\tau}{\longrightarrow} Q$}
\DisplayProof\quad\quad
\AxiomC{\emph{M is destructor-free}}
\RightLabel{\small{(Out)}}
\UnaryInfC{$ n\langle M \rangle.P \stackrel{n\langle M \rangle}{\longrightarrow} P$}
\DisplayProof \quad\quad
\AxiomC{\emph{M is destructor-free}}
\RightLabel{\small{(Inp)}}
\UnaryInfC{$ n(x).P \stackrel{n(M)}{\longrightarrow} P$}
\DisplayProof
$$
$$
\AxiomC{$M \in \mathfrak{F}(gt(P) \cup \bar{s})\;\;\;\bar{s} \mbox{\emph{ fresh}}$}
\RightLabel{\small{(AttackerOut)}}
\UnaryInfC{$n\langle * \rangle.P \stackrel{\boldsymbol{\nu}\bar{s}.n\langle M \rangle}{\longrightarrow} P$}
\DisplayProof \quad\quad
\AxiomC{$P \stackrel{\alpha}{\longrightarrow} Q $}
\AxiomC{$\forall n \in \bar{u}$: n $\not\in$ $names(\alpha)$}
\RightLabel{\small{(Res)}}
\BinaryInfC{$(\boldsymbol{\nu}\bar{u})P \stackrel{\alpha}{\longrightarrow} (\boldsymbol{\nu}\bar{u})Q$}
\DisplayProof
$$
$$
\infer[(BoundOut)]{(\boldsymbol{\nu}\bar{u})P
  \stackrel{\boldsymbol{\nu}\bar{s}.n\langle M
    \rangle}{\longrightarrow} (\boldsymbol{\nu}\bar{u'}) P'}
{P \stackrel{n\langle M \rangle}{\longrightarrow} P' & 
 \bar{s} \subseteq names(M)$ and $\bar{s} \subseteq \bar{u} &
 \bar{u'} = \bar{u} \setminus \bar{s}}
\quad\quad
\infer[(Cong)]
{P \stackrel{\alpha}{\longrightarrow} Q}
{P \equiv P' & P' \stackrel{\alpha}{\longrightarrow} Q' &
 Q' \equiv Q}
$$
}
\vspace{-0.2in}
\caption{Labelled Transition Semantics}\label{fig:lts}
\vspace{-0.25in}
\end{center}
\end{figure}\end{definition}
Our labelled semantics is not intended to characterize a complete notion of
behavioral equivalence as could be expected, but rather to allow the
observation of actions in our logic.  Despite not belonging to the
scope of this work, we can point out that our labelled semantics do not allow for a
complete characterization of behavioral equivalence, in the sense that
our rules reveal information in a way that induces a higher
discriminative power then that of behavioral equivalence.

\section{Logic}\label{sect:logic} 

Considering it is common to reason about security by reasoning about 
the knowledge of principals, we explore key aspects of 
dynamic spatial logics, such as local reasoning, 
to develop a logic that can reason about epistemic, 
dynamic and spatial properties of agents.

We propose an extension to a dynamic spatial logic \cite{Caires04:0} 
to enable reasoning at the term level. Our extension consists of adding two 
epistemic modalities: $\knows \phi$ denotes the ability of an agent 
to derive $\phi$ from its knowledge, and $\secret x.A$ allows us to mention
values that are only known by an agent (e.g. secrets).
Our intent is to couple the ability to reason about properties of space and behavior 
with that of reasoning about derivable information modulo the
equational theory. Our notion of knowledge is therefore the ability of
an agent to derive terms from the information it possesses.

\subsection{Syntax and Semantics}
The syntax and semantics of our logic are presented in Fig. \ref{fig:log_ss}. 
\begin{figure}[tp]
\begin{center}
{ \small
$$
\begin{array}{llll}
A,B & \Coloneqq & \mathbf{T} & \mbox{(True)} \\
   & \;\;\;\origbar & \neg A & \mbox{(Negation)}\\
   & \;\;\;\origbar & A \wedge B & \mbox{(Conjunction)}\\
   & \;\;\;\origbar & \mathbf{0} & \mbox{(Void)}\\
   & \;\;\;\origbar & A\;\origbar\; B & \mbox{(Composition)}\\
   & \;\;\;\origbar & \mathsf{H}x.A & \mbox{(Hidden quantification)}\\
   & \;\;\;\origbar & \alpha.A & \mbox{(Action)}\\
   & \;\;\;\origbar & \Box A & \mbox{(Always)}\\
   & \;\;\;\origbar & \Diamond A & \mbox{(Eventually)}\\
   & \;\;\;\origbar & \mbox{@}n & \mbox{(Free-name Predicate)}\\
   & \;\;\;\origbar & \knows \varphi & \mbox{(Knowledge)}\\
   & \;\;\;\origbar & \secret x.A & \mbox{(Secret quantification)}\\\\
\phi,\psi & \Coloneqq & \varphi \wedge \psi & \mbox{(Conjunction)} \\
   & \;\;\;\origbar & t & \mbox{(Term)}\\
   & \;\;\;\origbar & \top & \mbox{(True)}
\end{array}\quad
\begin{array}{lll}
P\models\mathbf{T} & \triangleq & \mbox{\emph {True}}\\
P\models\neg A &\triangleq & \mbox{\emph {not} }P\models A\\
P\models A \wedge B & \triangleq& P\models A \mbox{ \emph{and} } P\models B\\
P\models\mathbf{0} & \triangleq& P\equiv \mathbf{0}\\
P\models A\;\origbar\; B &\triangleq & \exists Q,R. P \equiv Q\;\origbar\; R\mbox{ \emph{and} }Q\models A \mbox{ \emph{and} } R\models A\\
P\models\mathsf{H}x.A & \triangleq&\exists Q.P\equiv(\boldsymbol{\nu}n)Q \mbox{ \emph{and} } Q\models A\{x \leftarrow n\}  \\
P\models\alpha.A &\triangleq &\exists Q.P\stackrel{\alpha}{\rightarrow}Q \mbox{ \emph{and} } Q\models A\\
P\models\Box A &\triangleq & \forall Q \mbox{ s.t } P\stackrel{\tau^*}{\rightarrow}Q \mbox{ \emph{then} } Q \models A\\
P\models\Diamond A &\triangleq & \exists Q.P\stackrel{\tau^*}{\rightarrow}Q \mbox{ \emph{and} } Q\models A \\
P\models \mbox{@}n & \triangleq & n\in fn(P)\\
P\models\knows \phi &\triangleq & P\vdash_k \psi \mbox{ \emph{and} } \psi \models \phi\\
P\models\secret x.A &\triangleq & \exists Q,t. P \equiv (\boldsymbol{\nu}k)Q \mbox{ \emph{and} } Q \models A\{x \leftarrow t \}\\
& & \mbox{ \emph{and} } Q \vdash_{k} \phi \mbox{ \emph{such that} } t \in \phi\\
& & \mbox{ \emph{and} } k \in names(t) 
\end{array}   
$$}
\vspace{-0.2in}
\caption{Logic Syntax and Semantics}\label{fig:log_ss}
\vspace{-0.3in}
\end{center}
\end{figure}
We refer to $\phi,\psi$ as knowledge formulas and ambivalently use $\phi,\psi$ to denote both knowledge formulas and finite sets of terms.
The boolean connectives are standard. $\mathbf{0}$ denotes the empty process; $A\;\origbar\; B$ denotes a process that
can be partitioned in two components,
one satisfying $A$ and the other satisfying $B$; $\mathsf{H}x.A$ allows us to mention restricted names of processes in formulas; 
$\alpha.A$ denotes a process can perform action $\alpha$
and continue as a process satisfying $A$; 
$\Box A$ and $\Diamond A$ denote ``always in  the future'' and ``sometime in the future'', respectively. 
$\knows \phi$ holds of a process that has the ability to derive the terms denoted by $\phi$, that is, the ability to know $\phi$; 
$\secret x.A$ holds of a process that satisfies property $A$ that
depends on a value that is secret to a process -- a term containing a
restricted name. It is also useful to define an auxiliary \emph{counting
predicate} (written as $\mathbf{n}$, where $n$ is a natural number), that allows us to count the number of sub-processes within a
process. For instance, a process consisting of a single thread would
satisfy the formula $\mathbf{1}$ defined as $\neg \mathbf{0} \wedge
\neg ( \neg \mathbf{0} \;\origbar\; \neg\mathbf{0})$, while a process
consisting of two sub-processes would satisfy the formula $\mathbf{2}$
defined as $\neg \mathbf{0} \wedge \neg\mathbf{1} \wedge \neg(\neg \mathbf{0} \;\origbar\;
\neg\mathbf{0} \;\origbar\; \neg\mathbf{0})$, and so on.

With this logic, we can state properties about the knowledge of
agents (and not only adversarial ones) over time,
such as ``it is never the case that the secret key is known by 3 subsystems'': 
$$\neg\Diamond\mathsf{H}\;key.(\knows\; key \;\origbar\; \knows\; key \;\origbar\; \knows\; key)$$ 
or ``it is always the case that 2 agents know the key and one does not'': 
$\Box \mathsf{H}\;key.(\knows\; key \;\origbar\; \knows\; key
\;\origbar\; \neg\knows\;key)$. 
Since the semantics of our logic blur together processes that are
structurally congruent (e.g. $P \;\origbar\;Q$ and $Q\;\origbar\;P$), 
we can use the free-name predicate to ``tag'' specific subsystems and
reason about their knowledge explicitly:
$\Box \mathsf{H}\;key.(\mbox{@}tag \wedge \knows\; key \;\origbar\; \mathbf{T})$ which denotes ``it is always the case 
that an agent with the free name \emph{tag} knows the key'' (this subsumes the need for an indexed knowledge 
operator such as that in \cite{mardarepriami06:0}).

Notice how the expressiveness of the logic arises from the ability
to combine the three types of modalities: dynamic ($\Box, \Diamond$), spatial ($\mathsf{H},\; \origbar\;$) 
and epistemic ($\knows$). 
The dynamic connectives allow us to range over a specific execution or all possible executions, 
the spatial connectives allow us to mention restricted names (usually
used to model keys and nonces) and to refer to subsystems, 
and the epistemic connectives allow us to analyze derivable terms of a process. 

The semantics for $\knows \phi$ pose a challenge in the sense that 
they use the notion of knowledge derivation from Section \ref{sect:terms}. 
While this definition is adequate from a semantic perspective, it makes use 
of the DY equational closure of a set which is not stable by reduction of terms, 
and thus doesn't provide a clear way of algorithmically determining if $\psi \models \phi$. 
We approach the problem with a purely logical approach 
and characterize knowledge derivation with a structural proof system for knowledge formulas, 
unlike the  approach of \cite{AbadiCortier06:0}.

\subsection{Proof System for Knowledge Formulas}\label{sect:proof}

Our proof system, formulated as a sequent calculus, is equipped with rules from the equational theory 
in order to consider the ability to combine terms to generate new information. Each rule of our calculus represents a 
possible computational step that an agent can perform on terms to produce a new term.
Intuitively, if a sequent $\Gamma \vdash \phi$ is derivable, the knowledge formula $\phi$ is deducible from the 
knowledge represented by $\Gamma$.
\begin{definition}[Proof System K for Knowledge Formulas]
The sequent calculus formulation of our proof system $K$ for knowledge formulas is defined by the rules of Fig.~\ref{fig:sequent}.
\begin{figure}[tp]
\begin{center}
{\small
$$\displaystyle
\frac{}{\;\; \Gamma ,A \vdash A\;\;}\mbox{\small{(Id)}} \;\;\;\;\;\;\;
\frac{\Gamma ,A,B \vdash C}{\;\;\Gamma , A \wedge B \vdash C\;\;}\mbox{ \small{($\wedge$: left)}} \;\;\;\;\;\;\;    \frac{\Gamma \vdash A \;\;\;\; \Gamma \vdash B}{\;\;\Gamma \vdash A\wedge B\;\;} \mbox{ \small{($\wedge$: right)}}
$$
For every constructor function symbol $f$ with arity n, such that $f \in \Sigma$:
\vspace{-0.05in}
$$\displaystyle
\frac{\Gamma \vdash t_{1} \dots \Gamma \vdash t_{n}}{\;\; \Gamma \vdash f(t_{1},\dots,t_{n})\;\;}\mbox{ \small{(funRight)}} \;\;\;\;\;\;\; \frac{\Gamma,f(t_{1}, \dots ,  t_{n}) \vdash C }{\;\; \Gamma , t_{1}, \dots ,  t_{n} \vdash C  \;\;}\mbox{ \small{(AttLeft)}}
$$
For every equation $f(t_{1}, \dots , t_{n}) = s \in E$:
\vspace{-0.05in}
$$\displaystyle
\frac{\Gamma, s \vdash C}{\;\; \Gamma,f(t_{1},\dots , t_{n}) \vdash C \;\;}\mbox{ \small{(DestrLeft)}}
$$}
\vspace{-0.25in}
\caption{Proof System for Knowledge Formulas}\label{fig:sequent}
\vspace{-0.25in}
\end{center}
\end{figure}
\end{definition}
The rules for identity and conjunction are standard. Rule
\emph{funRight} states that we are justified in concluding a complex
term if we can derive its subterms. 
Rule \emph{AttLeft} states that all that can be derived from a complex
term $f(t_{1}, \dots ,  t_{n})$ 
can also be derived from its subterms; rule \emph{DestrLeft} reflects the equalities of the equational theory: 
what can be deduced
from $s$ can also be deduced from terms equal to $s$ under the equational theory. 

For the sequent calculus $K$, we establish the results of \emph{soundness}, \emph{completeness} and \emph{decidability}. 
\begin{theorem}[Soundness of K]
Given a set of terms S and a term $A$, if $S \vdash A$ then $S \models A$.\\
Proof: By induction on the derivation of $S \vdash A$\qed
\end{theorem}
\begin{theorem}[Completeness of K]\label{theor:completeness}
Given a set of terms S and a term $A$, if $S \models A$ then $S \vdash A$.
\end{theorem}
\begin{theorem}[Decidability of K]\label{theor:decidability}
For any set of terms S and term $A$, $S \vdash A$ is decidable.
\end{theorem}

The proofs of completeness and decidability rely on a finite approximation result for
the DY equational closure of a set of terms. More concretely, for each finite
set of terms $S$ and equational theory, it is possible to build a finite set $b(S)$ from which
all terms in the DY equational closure of $S$ may be determined.
\begin{proposition}[Approximation of $\mathfrak{F}(S)$]
Let $S$ be a finite set of terms. We may construct in polynomial time
an approximation to $\mathfrak{F}(S)$, named $b(S)$, a finite set with the following property:
$$
\forall M \in \mathfrak{F}(S) , \exists \;C, \bar{t} \in b(S) \mbox{ such that } M = C[\bar{t}]
$$
where $C[-]$ is a functional context solely built out of constructors.\\
Proof: 
The finite approximation $b(S)$ is built from the terms of $S$ by interpreting the 
rewrite rules of the theory as contexts of a bounded size. 
Therefore, applying 
function symbols to terms of $S$ up to the bound of the context 
produces a new term by then applying the rewrite rule. 
This procedure is iterated, eventually reaching a fix-point, due to the subterm
convergency property of the equational theories (the idea is that each time 
we produce a new term, the term will be smaller then the terms
used to generate it). The resulting computable set has the 
property that defines our approximation \cite{toninhocaires09:0}. \qed
\end{proposition}

The approximation $b(S)$ is such that all terms of $\mathfrak{F}(S)$ can be built 
from terms of $b(S)$ just by applying constructors, 
no longer requiring the equations from the theory.
Completeness follows from the fact that our proof system is able 
to emulate the computation steps required to generate the approximation. 
Given a set of terms $S$, $b(S)$ is generated
by applying functions to terms of $S$, applying a rewrite rule to the resulting term and iterating. 
Thus, our proof system is complete since the computation steps of $b(S)$ may be emulated
by the rules of proof system K, and we may then apply function symbols to terms of $b(S)$ to 
produce terms of $\mathfrak{F}(S)$. 
The latter is trivial due to \emph{funRight} and \emph{AttLeft}. The former we prove through the following lemma.
\begin{lemma}{\textbf{(Completeness of K i.r.t the Approximation).}}\label{lemma:approx}
Given a set of terms $S$,if $t \in b(S)$ then $S \vdash t$.\\
Proof: 
Through instances of \emph{AttLeft} it is possible to apply 
functions to terms of $S$ up to the bound of the context used in $b(S)$. 
Through an instance of \emph{DestrLeft} the corresponding rewrite rule can applied,
and through \emph{Id} the new term is derived at the root of the proof tree \cite{toninhocaires09:0}.\qed
\end{lemma}
To emulate the iteration with the proof system, that is, to perform similar computations with $S$
and the new term, the auxiliary result of reasoning with cuts is required.
\begin{lemma}{\textbf{(Cut Admissibility in K).}}
If $\Gamma \vdash A$ and $\Gamma, A \vdash C$ then $\Gamma \vdash C$.\\
Proof: See \cite{toninhocaires09:0}.
\end{lemma}
Using Cut, the proof system is able to emulate the iterative procedure
by building the previously described proof tree that allows the derivation of a new term, 
and using the new term as the cut formula. This technique can then be applied
to produce any term of $b(S)$, as required. 
Since the computation of $b(S)$ always terminates, Theorem \ref{theor:decidability} holds.

\subsection{Model-Checking}

We know that model-checking is decidable for the logic without the new modalities \cite{Caires04:0}, 
for the class of bounded processes. 
Therefore, we need only show that our two modalities preserve decidability.
\begin{proposition}[Decidability of model-checking $\knows$]
Let $\phi$ be a finite set of terms. Checking that $P \models \knows \phi$
is decidable.
\end{proposition}
The above proposition holds since for any process $P$ it is possible to collect its set of relevant terms $\psi$ ($P \vdash_k \psi$),
compute the finite approximation $b(\psi)$ and check that each term in $\phi$ can be constructed from terms of $b(\psi)$ by application
of constructors.
\begin{proposition}[Decidability of model-checking $\secret x.A$]
Checking that $P \models \secret x.A$ is decidable.
\end{proposition}
Decidability of $\secret x.A$ follows from the fact that if $P \equiv (\boldsymbol{\nu}n)Q$, it is possible to collect the set $\psi$ of relevant terms of process
Q, pick some term $t$ from $\psi$ that contains the name $n$ and check that $Q\models A\{x \leftarrow t\}$. 
Given that model-checking the core logic with $\knows$ is decidable, it follows that checking $P \models \secret x.A$ is decidable and
therefore model-checking for our logic is decidable for the class of
bounded processes.

\begin{theorem}[Decidability of Model-Checking]
Checking that $P \models A$ is decidable for the class of bounded processes.
\end{theorem}

\section{Expressiveness and Extensions}

Having presented our framework, we discuss some extensions to our work
that can be used to model and analyze
systems. In particular, we discuss the representation of attackers
and modeling and verification of correspondence assertions
\cite{884188} in our framework.

\subsection{Modeling Attackers}\label{sect:attackers}

To analyze a security protocol one usually needs to consider how it
behaves in any possible environment. While our logic focuses on the
analysis of closed systems, it is possible to verify properties
of a system in an arbitrary environment, by 
internalizing an arbitrary attacker in the system. The general idea
is that, for any process $P$, we may determine a process $Q$ (making
essential use of the attacker prefix construct) such that $P
\origbar Q$ reaches some state whenever $P$ reaches an equivalent
state when placed in an arbitrary environment.
While the explicit specification of an attacker for a given
protocol may not be easy, our approach to
represent the attacking environment is quite different and general,
and may indeed be used to find attacks (see example in
Section \ref{sect:correspondence}).  
We can generically model
a Dolev-Yao \cite{891726} attacker in our framework by considering the
number of message exchanges and the communication channels used in a
protocol.

Considering an arbitrary protocol modeled as a process, the role of an
attacker is to intercept all communications of the principals and be
able to inject any message it can produce, given its knowledge at the
time, at any point where a principal expects to receive a message
(cf. our attacker output). Thus, a Dolev-Yao attacker consists of
a process that for all outputs of the protocol performs an input
(storing the received message) and for all inputs performs an attacker
output.  For instance, consider the following protocol, where $K$ is a
shared key, $N$ a fresh value and $K_{ab}$ a session key generated by
$A$:
$$
\begin{array}{l}
A \rightarrow B :  \{ K_{ab},N \}_K \\
B \rightarrow A : \{ N-1 \}_{K_{ab}}
\end{array}
$$
In our process model, such a protocol would be represented as done in Fig. \ref{fig:example_model} (we omit the signature and equational theory).
\begin{figure}[tp]
\begin{center}
{\small $$
\begin{array}{lll}
A(K) & \triangleq & (\boldsymbol{\nu}K_{ab},N)\; c\langle \mathtt{enc}(\mathtt{pair}(K_{ab},N),K)\rangle.c(x).[N-1 = \mathtt{dec}(x,K_{ab})]\\
B(K) & \triangleq & c(x).\mathtt{let}\; K_{ab} = \mathtt{fst}(\mathtt{dec}(x,K))\;N = \mathtt{snd}(\mathtt{dec}(x,K))\;\mathtt{in}\; c\langle \mathtt{enc}(N-1,K_{ab})\rangle\\
Sys & \triangleq & (\boldsymbol{\nu}K)\; (A(K)\:\origbar\;B(K))
\end{array}
$$}
\vspace{-0.25in}
\caption{Modeling the Example}\label{fig:example_model}
\vspace{-0.3in}
\end{center}
\end{figure}
An attacker for this protocol, following our \emph{attacker schema} is presented in Fig. \ref{fig:example_att}.
We can then state that it is never the case that the attacker can know one of the keys used in the protocol.
\begin{figure}[tp]
\begin{center}
$$
\begin{array}{lll}
Attacker & \triangleq & c(x).c\langle*\rangle.c(y).c\langle*\rangle.mem\langle x,y\rangle\\
World & \triangleq & (Sys \;\origbar\; Attacker)\\
World & \models & \neg \Diamond \mathsf{H}k.(\mathbf{2} \;\origbar\; (@mem \;\wedge\;\knows k))
\end{array}
$$
\vspace{-0.25in}
\caption{An Attacker for the Example}\label{fig:example_att}
\vspace{-0.3in}
\end{center}
\end{figure}
While some minor effort of representing
an attacker is necessary, we can easily represent a generic attacker
for a protocol by following a pre-determined schema.  

We currently only consider finite protocols, modeled as processes in our calculus that use a
communication channel $c$ as their communication medium (written
$P_{c}$). We have not pursued infinite protocols as of yet, but we
believe it to possible to extend our approach to infinite protocols by
defining the attacker as a recursive process with a parallel store
(that is used to store the messages of the protocol). To analyze such
a system, we would then employ recursive formulas by using the
fixpoint operators of the logic.

Our attacker for finite protocols is defined as follows: For each output on $c$,
the attacker performs an input on $c$ (and stores the message). For
each input on $c$, the attacker performs an attacker output.

\begin{definition}[Attacker Generation Procedure]
Given a process $P_c$ that models a finite protocol, the set $S$ that tracks the attacker memory, an attacker for $P$
can be generated by procedure $|Attacker|(P,S)$ defined in Fig. \ref{fig:attacker_gen} ($x$ and $m$ are fresh in $P$ and the attacker).
\begin{figure}[tp]
\begin{center}
{\small \begin{program}
\PROC |Attacker|(P,S) \BODY
\IF P\stackrel{\alpha}{\rightarrow}Q \AND \alpha = \mbox{input on } c \THEN c\langle*\rangle.|Attacker|(Q,S)  \FI	
\IF P\stackrel{\alpha}{\rightarrow}Q \AND \alpha = \mbox{output on } c \THEN c(x).|Attacker|(Q,S \cup x)  \FI
\IF P\stackrel{\alpha}{\not\rightarrow} \THEN m\langle x_1 , \dots , x_n \rangle \mbox{ where } x_i \in S \FI\ENDPROC
\end{program}}
\vspace{-0.1in}
\caption{Attacker Generation}\label{fig:attacker_gen}
\vspace{-0.3in}
\end{center}
\end{figure}
\end{definition}
The generation procedure produces the necessary actions by inspection
of the process dynamics: if an output can occur in the process, the
attacker intercepts the message and memorizes it; if an input can
occur in the process, the attacker injects any message it can produce
from its knowledge; in the case where the protocol has no more
actions, we represent the attacker memory with an output $m\langle x_1
, \dots , x_n \rangle$, modeling the attacker's memory throughout the
protocol run. We thus show how an attacker can be extracted by
inspection of the considered protocol.
We can show that this attacker is general in our framework, in the
sense that it can simulate the behavior expected from an adversarial
medium (c.f. Dolev-Yao attacker). Note that this result does not yet
fully apply to our tool implementation, as we discuss later in this
section.

\begin{definition}[K-Set]
Given a process $P$, we define its K-Set $K_{P}$, the set of all terms known by $P$ as:
$
K_{P} \triangleq \{ t \;\origbar\; P \models \knows \;t\; \}
$
\end{definition}

\begin{theorem}[Monotonicity of K-Sets under Synchronization]\label{theor:monoton}
Let $P_c$ and $A$ be a processes such that $P_c \stackrel{c\langle M\rangle}{\longrightarrow} P'_c$ and $A \stackrel{c(x)}{\longrightarrow} A'$.
We have that $K_{A'} \subseteq \mathfrak{F}(K_{A} \cup M)$.
\end{theorem}

We begin with the K-Set of a process, the set of all terms known by
the process that we observe in our logic, and we show that the
evolution of arbitrary processes' K-Sets through synchronization is
monotonic: the resulting process' knowledge will be
a subset of the initial process' knowledge, plus any received
messages.  We state a similar property of our generated attacker's
K-Set. Over time, the attacker's K-Set captures all messages exchanged
in the protocol.

\begin{theorem}[Monotonicity of Attacker Storage]\label{theor:attacker_stor}
Let $P_c$ and $At$ be processes such that\\ $At = |Attacker|(P_c,\{\},c)$,
$P \stackrel{c\langle M\rangle}{\longrightarrow} P'$ and $At \stackrel{c(x)}{\longrightarrow} At'$.
We have that $K_{At'} = \mathfrak{F}(K_{At} \cup M)$.
\end{theorem}

Our Attacker Simulation (Lemma \ref{theor:as}) and Process Knowledge
(Lemma \ref{theor:pk}) lemmas provide some insight on the
expressiveness of our attacker. Lemma \ref{theor:as} shows
that a generated attacker, can obtain as much knowledge as an
arbitrary process interacting with a finite protocol. 
Lemma \ref{theor:pk} states a similar
property, regarding the knowledge a finite protocol may obtain while
interacting with our attacker.

\begin{lemma}[Attacker Simulation]\label{theor:as}
Let $P_c$ and $A$ be processes.\\
If $(\boldsymbol{\nu}\bar{n}) (P_c \;\origbar\; A) \longrightarrow (\boldsymbol{\nu}\bar{n}) (P'_c \;\origbar\;A')$ and $At = |Attacker|(P_c,S)$ with $K_A \subseteq K_{At}$
then $\exists At', S'$ such that\\ 
$(\boldsymbol{\nu}\bar{n})(P_c \;\origbar\;At )\stackrel{*}{\longrightarrow} (\boldsymbol{\nu}\bar{n})(P'_c \;\origbar\;At')$ and $At' =  |Attacker|(P'_c,S \cup S')$ and
$K_{A'} \subseteq K_{At'}$.
\end{lemma}

\begin{lemma}[Process Knowledge]\label{theor:pk}
Let $P_c$, $A$ be processes and $\phi$ a knowledge formula.\\
If $(\boldsymbol{\nu}\bar{n})(P_c \;\origbar\; A)
\stackrel{*}{\longrightarrow}  (\boldsymbol{\nu}\bar{n})(P'_c \;\origbar\; A') \mbox{ and } P'_c \models \knows \phi$ and $At = |Attacker|(P_c,S)$ with $K_A \subseteq K_{At}$
then $\exists At', S'$ such that
$(\boldsymbol{\nu}\bar{n})(P_c \;\origbar\; At) \stackrel{*}{\longrightarrow} (\boldsymbol{\nu}\bar{n})(P'_c\;\origbar\; At') \mbox{ and } P'_c \models \knows \phi$ and $At' = |Attacker|(P'_c, S \cup S')$.
\end{lemma}

Furthermore, from Lemma \ref{theor:pk} follows that, in our logic, a
finite protocol interacting with an arbitrary process is
indistinguishable from one interacting with our attacker.  
Combining these results, we can show that our attacker can behave as one
would expect of an adversarial Dolev-Yao agent.
\begin{theorem}[Preservation of Satisfaction]
  Let $P_c$ and $A$ be processes and $A$ any formula.  If\\
  $(\boldsymbol{\nu}\bar{n})(P_c \;\origbar\; A) \stackrel{*}{\longrightarrow} 
  (\boldsymbol{\nu}\bar{n})(P'_c \;\origbar\; A') \mbox{ and } P'_c
  \models A$ and $At = |Attacker|(P_c,S)$ with $K_A \subseteq K_{At}$
  then $\exists At', S'$ such that
  $(\boldsymbol{\nu}\bar{n})(P_c \;\origbar\; At) \stackrel{*}{\longrightarrow} (\boldsymbol{\nu}\bar{n})(P'_c\;\origbar\; At') \mbox{ and } P'_c \models A$ and $At' = |Attacker|(P'_c, S \cup S')$.
\end{theorem}
Notice that this result follows from the fact that message
size for the attacker output prefix is unbounded. Our
implementation currently bounds the generated message, to ensure tractability,
and thus sacrificing completeness. However, as shown
in \cite{Rus:Tur:03}, it is possible to compute a finite bound on the
message size required to find an attack. The implementation
of this result we leave for future work.
It is anyway important to note that our method is already \emph{sound and complete}
for passive attackers, even for the case of non finite processes
(eg. we may consider any finite control system, or bounded in the
sense of \cite{Caires04:0}).


\subsection{Modeling Correspondence Assertions}\label{sect:correspondence}
Correspondence assertions are a technique for verifying authentication
properties in protocols \cite{884188}.  The idea is that the
model of each principal in a protocol is refined with begin/end
events, named \emph{correspondence assertions}, at each stage of an
authentication procedure.  Authentication will be established if, for
every run of the protocol, all end events for each stage are preceded
by a matching begin event. To illustrate the idea, consider the following protocol:
{\small
$$
\begin{array}{l}
A \rightarrow B :  \{ N \}_k ;\qquad  B \mbox{ asserts the reception of
} N \\
B \rightarrow A : \{ h(N) \}_k ;\quad A \mbox{ asserts the reception of } h(N)  \\
\end{array}
$$}
Principals $A$ and $B$ share a symmetric key $k$,
$N$ is a fresh value and $h$ is a one-way hash function. 
When B receives $\{ N \}_k$ it asserts the beginning of the run of the
protocol. $B$ sends message $\{ h(N) \}_k$ so that $A$ can verify the
freshness of the run, by comparing the received value with its own
hash of $N$. If the test succeeds, $A$ asserts the reception of $h(N)$
and the end of the run.  To check correspondence, one has to check
that every run of the protocol, in the presence of an adversary, would
be such that $A$'s end assertion is always preceded by $B$'s begin
assertion, that is, $A$ only ends if $B$ was involved in the protocol.

Using our framework, we can model correspondence assertions by
representing the assertion as an output on a channel that is
irrelevant to the protocol, and then observing the existence of such
outputs with our logic. For instance, our example could be modeled as
done in Fig. \ref{fig:correspondence} (note the \verb=attacker_depth=
parameter set to 2 due to the size of the second message). We can also
successfully handle the case where we consider a faulty system that
leaks $k$ to the attacker (and thus correspondence does not hold), as
presented in Fig. \ref{fig:not_corresp}. 
\begin{figure}[tp]
\begin{center}
{\scriptsize
\begin{verbatim}
parameter attacker_depth = 2;

defproc A(k) = new N in c!(enc(N,k)).c?(x).[dec(x,k)=h(N)].end!(h(N));
defproc B(k) = c?(x).(begin!(dec(x,k)) | c!(enc(h(dec(x,k)),k)));
defproc Sys = new k in (A(k) | B(k));
defproc Attacker = c?(v).c!(*).s!(v);
defproc World = (Sys | Attacker);

defprop begin = <begin!> true;
defprop end = <end!> true;
defprop corrsp = always (end => begin);
check World |= corrsp;
Processing...
* Process World satisfies the formula corrsp *
\end{verbatim}
}
\vspace{-0.1in}
\caption{Checking Correspondence in a Toy Protocol}\label{fig:correspondence}
\vspace{-0.25in}
\end{center}
\end{figure}
\begin{figure}[tp]
\begin{center}
{\scriptsize
\begin{verbatim}
...
defproc Sys = new k in (c!(k).(A(k) | B(k)));
defproc Attacker = c?(u).c?(v).c!(*).s!(v,u);
defproc World = (Sys | Attacker);
...
check World |= corrsp;
Processing...
* Process World does not satisfy the formula corrsp *
\end{verbatim}
}
\vspace{-0.1in}
\caption{Checking Correspondence in a Broken Toy Protocol}\label{fig:not_corresp}
\vspace{-0.3in}
\end{center}
\end{figure}


\section{Concluding Remarks and Related Work}\label{sect:conc} 

In this paper we have introduced a dynamic spatial epistemic logic for 
a variant of the applied $\pi$-calculus aimed at reasoning about security protocols.
We explore the application of spatial and epistemic reasoning to the 
several agents involved in a security protocol, be they principals or adversaries.
In our work, we can reason about the knowledge of the several agents of a protocol 
and how it can evolve over time.
Model-checking for the logic is shown to be decidable for an interesting class 
of processes and cryptographic primitives. 

Our framework allows an interesting degree of freedom in the analyses it can perform, 
not only allowing one to reason directly about knowledge of principals
and attackers but also
enabling reasoning with correspondence assertions, which is an important addition 
to the range of available techniques. 
Moreover,
our internalization of attackers, which does not require a 
complete behavioral specification, is able to accurately emulate the 
behavior of a Dolev-Yao attacker, enabling reasoning about the dynamics
and knowledge of such an attacker.

Finally, the decidability result for our logic allowed us to implement a 
model-checking algorithm as a proof of concept extension to the SLMC tool.
The main difference between the tool and the theory is that our attacker 
outputs are parametrized with a maximum message size, to
bound the state space. 
This is the main limitation of the current version
of our tool, since it does not yet fully capture the expressiveness of our
attacker modeling, given that our results employ a more powerful version
of the attacker output.

Overall, we have produced an interesting framework for
protocol analysis, the first employing dynamic spatial logics.
enabling a very natural (yet precise) way of reasoning about security
protocols, all the while allowing reasoning with previously
established techniques. Note, however, that our tool is merely a proof
of concept of the developed framework, not aimed to compete 
with more mature tools for protocol analysis
such as Avispa \cite{Armando05}, Scyther \cite{Cr2008Scyther}, Casper
\cite{Lowe98casper:a} or ProVerif \cite{BlanchetCSFW01}. The main
point of divergence of our approach and the ones
mentioned before is that instead of mainly focusing on a set of built-in
properties, we focus on a generic property language (our logic)
and explore its expressiveness.

In terms of related logics, Kremer et al.~\cite{CDK-forte09} have
proposed an epistemic
logic for the applied $\pi$-calculus. However, their logic lacks the ability to
reason about spatial properties, which is a key element in allowing
reasoning about individual agents. Their epistemic modalities focus
solely on attacker knowledge, not allowing one to state a
property such as that of our introductory example where we care about
the knowledge of the attacker but also of the agents within the
system.

Another closely related logic is Datta et al.'s PCL
\cite{Datta07protocolcomposition}. PCL is a well established
protocol analysis logic that allows one to verify properties
of protocols modelled in a CCS style calculus by 
reasoning about events that occur in traces of the protocol run.
While we focus on the combined reasoning about knowledge and spatial
distribution of a
protocol, PCL is designed to reason about the composition of several
protocols and thus its analyses are more sophisticated than ours
(reasoning about invariants in the protocol composition interleavings).

Mardare and Priami have also proposed a dynamic epistemic spatial
logic \cite{mardarepriami06:0} without the
issues of security in mind.  Their logic is hence substantially
different from ours, interpreting knowledge as the possibility
of observing actions of other processes and not as the ability
to know some piece of information. Being based on CCS,  such an approach
is not suitable for reasoning about the flow of messages within a system,
which is one of our main goals.

For future work we wish to further study the problem of attacker
representations, aiming at an expressiveness result along the lines of
Theorem~\ref{theor:pk} that does not require the attacker to be able
to produce a message of an arbitrary size (this should follow from the
result of \cite{Rus:Tur:03}). This result will be key in 
removing the previously discussed limitation of our tool.
\paragraph{Acknowledgments}
The first author acknowledges support for this research provided by
the Fundação para a Ciência e a Tecnologia
through the Carnegie Mellon Portugal Program under Grant SFRH / BD / 33763 / 2009.
We  thank Mário Pires and Pedro Adão for their
interesting comments on some version of this work. We also acknowledge
the anonymous reviewers for their comments and suggestions.

\bibliographystyle{eptcs}
\bibliography{bib}

\end{document}